\documentclass[a4paper,%
   a4paper,%
   BCOR12mm,%
   12pt,%
   abstracton,%
   pointednumbers,%
   tablecaptionabove,%
   footinclude,%
   halfparskip,%
   ]{scrartcl}
\usepackage{epsfig}
\usepackage[square,comma,numbers,sort&compress]{natbib}
\newcommand{\M}{\mathcal{M}} %matrix
\newcommand{\G}{\mathcal{G}} %graph
\newcommand{\V}{\mathcal{V}} %set of vertices
\newcommand{\E}{\mathcal{E}} %set of edges
\newcommand{\C}{\mathcal{C}} %cluster
\newcommand{\A}{\mathcal{A}} %adjacency matrix
\newcommand{\degree}{\Delta} %vertex degree
\newcommand{\power}{\varphi} %power law
\newcommand{\SN}{\mathcal{SN}} %social network
\newcommand{\PU}{\mathcal{P}}%processing unit
\begin{document}
\title{A Hybrid Graph-drawing Algorithm for\\Large, Naturally-clustered, Disconnected Graphs}
%\numberofauthors{2}
\author{Toni-Jan Keith P. Monserrat, Jaderick P. Pabico and Eliezer A. Albacea
%\alignauthor Toni-Jan Keith P. Monserrat\titlenote{Instructor}\\
   %\affaddr{Institute of Computer Science}\\
   %\affaddr{\UPLB}\\
   %\affaddr{College 4031, Laguna}\\
   %\email{tjmonserrat@uplb.edu.ph}
%\alignauthor Jaderick P. Pabico\titlenote{Associate Professor 6}\\   
   %\affaddr{Institute of Computer Science}\\
   %\affaddr{\UPLB}\\
   %\affaddr{College 4031, Laguna}\\
   %\email{jppabico@uplb.edu.ph}
}
\date{}
\maketitle
\begin{abstract}
In this paper, we present a hybrid graph-drawing algorithm (GDA) for layouting large, naturally-clustered, disconnected graphs. We called it a hybrid algorithm because it is an implementation of a series of already known graph-drawing and graph-theoretic procedures. We remedy in this hybrid the problematic nature of the current force-based GDA which has the inability to scale to large, naturally-clustered, and disconnected graphs. These kinds of graph usually model the complex inter-relationships among entities in social, biological, natural, and artificial networks. Obviously, the hybrid runs longer than the current GDAs. By using two extreme cases of graphs as inputs, we present in this paper the derivation of the time complexity of the hybrid which we found to be $O(|\V|^3)$. 
\end{abstract}
%\keywords{graph drawing, hybrid algorithm, large disconnected graph, clustered graph}

\section{Introduction}
% why is the problem interesting?
Information that abstractly describes the inter-relation-ships among entities in most complex systems is mathematically represented using graphs. Graphs as tools are an intuitive approach for visualizing entities because they make it easier for humans to understand the relationships between different entities. Because of this, graph visualizations of entities, as well as that of processed data, are used in many types of applications. For example, computer science concepts are usually easier to understand with the use of visualization concepts such as data flow diagrams, subroutine-call graphs, program nesting trees, object-oriented class hierarchies, entity-relationship diagrams, organization charts, circuit schematics, knowledge-representation diagrams, logic trees, and networks. Other fields of sciences also use graph visualization to represent information like concept lattices, evolutionary trees, molecular drawings, and maps and map schematics~\cite{DETT99}.

Because of the utility of graph visualization for viewing data that can be understood by the user in a vast number of applications, many  techniques were devised for drawing graphs efficiently and beautifully. Since the first paper by Knuth in 1963 on drawing flowcharts for visualization purposes~\cite{Knu63,DETT99}, there are now about 300 existing algorithms on graph drawing itself, some of these have improved the existing ones by utilizing the research advances made in topological and geometrical graph theory, graph algorithms, data structures, computational geometry, visual languages, graphical user interfaces and software visualization~\cite{DETT99}. However, given the numerous available algorithms, there is no one-size-fits-all graph drawing algorithm for any given graph. It is also important to identify the class to which a certain graph belongs. This is because several graph-drawing algorithms can only make effective visualizations on certain graph classes. Additionally, there are several approaches that exist in drawing graphs. Some of these approaches are drawing conventions, aesthetics, constraints and efficiency. These approaches include topology-shape-metrics, hierarchical, visibility, augmentation, divide and conquer, and force-directed.

In the current effort, we developed a hybrid force-directed approach algorithm based on the one developed by~\citet{KK89}. Here, we used a clustering algorithm called Markov cluster algorithm to cluster the original vertices into sub-graphs. We then used the original Kamada-Kawai (KK) force-directed algorithm to draw the vertices in each sub-graph. We considered each sub-graph as a big ``phantom'' vertex and applied the Iterative Kamada-Kawai (IKK) algorithm to draw the respective locations of the non-uniform-sized phantom vertices.

In this paper, we analyze the runtime of our hybrid graph drawing algorithm (HGDA). We illustrate our derivation by considering input graphs in extreme cases: a fully connected graph $\G_a(V_a,E_a)$ and a graph with no edges $\G_b(V_b,\emptyset)$. With these input graphs, we found out that HGDA has $O(|\V|^3)$ runtime complexity.

\section{Review}

Because graph drawings are used primarily to visualize information in a more understandable way, there are certain criteria that should be met when doing it. Drawing graphs should obviously include the type and properties of the graph to be drawn. This is important because several graph drawing algorithms are only designed to efficiently work on certain types of graphs. It is also essential to know that there is no optimum drawing for any graph because human perception changes from every individual. It should be noted that although the product of a graph-drawing algorithm may be subjective, it also has objective criteria such as drawing convention, aesthetic and constraints.

For a graph drawing to be admissible, it has to have some drawing conventions that it should follow. Examples of these conventions are having polyline for edges, using planar mathematics for layouting, and using grids to locate the vertices. A certain type of convention that is often used in graph drawing theories~\cite{DETT99} is the straight-line drawing. To objectively evaluate the aesthetics of a graph drawing, it specifies graphic properties of drawing that can achieve readability at the least. Some common aesthetic evaluation includes minimization of the total number of crossings between edges and minimization of the drawing area. These two efficiently use the drawing space without sacrificing the readability of the relationship between vertices~\cite{BFN85, PCJ96, STT81}. Additionally, constraints must also be considered specifically when drawing sub-graphs. Creating certain constraints on position and space provides how each subgraph should be drawn. Example of a common constraint would have a given vertex be drawn at the center of the drawing area. Another one is to have some of vertices be clustered or enclosed within a predefined shape~\cite{KMS94, TDB88}.

Because of these criteria, several approaches in graph drawing were established. One of these approaches  is through the use of force-directed algorithms (FDA). Due to their flexibility, ease of implementation and often-pleasant drawings, FDA are often used and improved~\cite{QE01}. Conventionally, FDA use straight-line drawings to draw edges in undirected graphs. FDA simulate some ``force'' that is directed to each vertex. When the minimal energy of the whole system is already achieved, the position of the vertices in the graph are said to be in its balanced state. To find the balanced state of the graph, FDA incorporate two main functions: (1) The  force model that simulates the forces acting on each of the vertex; and (2) An iterative algorithm to find the local minimal energy configuration~\cite{DETT99}. 

The KK algorithm takes in a connected graph $\G(\V, \E)$ and uses the graph theoretic distance (GTD) between each pair of vertices $u\in\V$ and $v\in\V$ as its force model. GTD between vertices~$u$ and~$v$ is calculated as the number of edges on a shortest path from vertex~$u$ to vertex~$v$. Usually, the aim of the FDA that uses GTD as a force model is to find the Euclidean distance between~$u$ and~$v$ to be approximately proportional to their GTD. KK includes an energy or spring view in the GTD~\cite{DETT99, KK89}. Because of this, KK was able to create symmetric drawings with relatively few edge crossings, which is practically similar to drawing isomorphic graphs~\cite{KK89}.  It should be noted, however, that KK only focused on fairly simple graphs. Originally, it was intended to solve undirected, non-weighted, simple and fully connected graphs~\cite{BM06}. An obvious problem for KK is the its inability to scale to handle large graphs. This inability is common also for other FDA. FADE~\cite{QE01}, a fast algorithm for two-dimensional drawing of large undirected graphs, was one of the more successful implementations of FDA that scale to larger graphs. It uses clustering before applying FDA, although primarily to lessen the computational time, and secondarily for maintaining the visualization better~\cite{QE01}.

There are many ways to cluster large graphs into manageable sub-graphs. Examples of these are the graph theoretic clustering~\cite{karypis98} and the geometric clustering~\cite{miller91} procedures like the ones being used in FADE, and the Markov Cluster Algorithm (MCL)~\cite{Dongen69}. One of the advantages of MCL is that it does not have any high level procedural rules for splitting or joining groups. The idea of MCL is to simulate a system of ``current'' $\C$ flowing inside the graph, promote that system when~$\C$ is strong, or demote the system when~$\C$ is weak. The computational paradigm is that~$\C$ between natural groups in the graph will wither away, revealing the cluster or sub-graph~\cite{Dongen69}.

Clustering a graph into sub-graphs defines the structure and natural clusters within the graph. By doing so, it arranges the vertices in the adjacency matrix~$\A$ by creating blocks of ``1s'' diagonally in~$\A$ where the clusters are formed. This makes it easy for the FDA to find the equilibrium by re-ordering the vertices according to their connections within and between the clusters, as opposed to the original procedure of randomly arranging vertices in~$\G$~\cite{Schaeffer07}.

\section{Theoretical Framework}

\subsection{Preliminary}

A graph~$\G$ is a pair of sets~$(\V,\E)$. $\V$ is the set of vertices and the number of vertices $n = |\V|$ is the order of the graph. The set $\E$ contains the edges of the graph. In an undirected graph, each edge is an unordered pair $\{v, w\}$. A vertex~$w$ is adjacent to a vertex~$v$ if and only if $(v, w)$ is an element of the set~$\E$. In an undirected graph, the abstract relationship represented by~$(v, w)$ is the same as that of~$(w, v)$.
 
A path in a graph is a sequence of vertices $w_1, w_2, \dots, w_n$ such that there exists an edge $(w_i, w_{i+1})$ where $1 \le i < n$. The length of the path is equal to number of edges $(n-1)$, where $n=|\V|$ is the number of vertices that runs along that selected path. 

A simple path is a path such that all vertices are distinct, with the exception of the first and last vertex of the path, which can be the same vertex~\cite{Weiss93}.

A graph $G'(V', E')$ is a sub-graph of $\G(\V,\E)$ if $V'\subset\V$ and $E'\subset\E\bigcap(V' \times\V)$.
 
A graph $\G(\V,\E)$ with $n=|\V|$ vertices can be described by an $n \times n$ adjacency matrix $\A$ whose rows and columns correspond to vertices. The matrix elements $A_{u,v}  = 1$ if $(u, v)$ is part of $\E$. $A_{u,v} = 0$ otherwise. A graph is connected if there is a path between~$u$ and~$v$ for each pair of vertices~$u$ and~$v$.

\subsection{Clustered and disconnected graphs}

\begin{figure}[htb]
\centering\epsfig{file=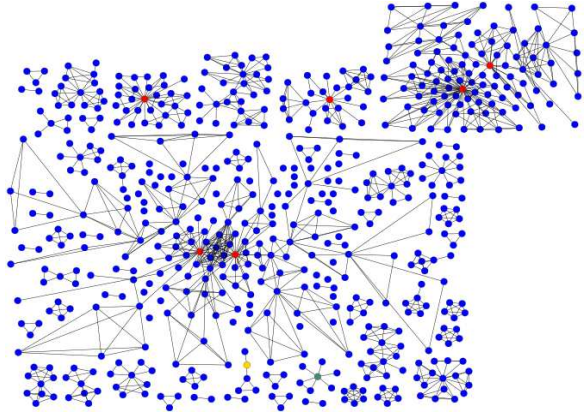, width=3in}
\caption{An example large, naturally-clustered and disconnected graph~$\G$ drawn using KK. This graph is based on the co-authorship network of Filipino computer scientists created by~\citet{pabico10} with $|\V|=605$. Notice that there exist some pronounced vertex clusterings in each connected component.}\label{fig:author-author}
\end{figure}

Graphs that are of small-world, scale-free characteristics are naturally clustered with some disconnected components. Small-world graphs are characterized by a very small network diameter, which usually values within six for naturally-occurring social networks $\SN$~\cite{travers69,watts98}. The degree~$\degree_i$ of a vertex $v_i$ counts the number of incident edges of~$v_i$. A symmetric matrix~$\A_{i,j}$ represents an undirected graph~$\G$, where $\A_{i,j}=\A_{j,i}=1$ if $v_i$ is incident to $v_j$. Thus, $\degree_i=\sum_{j=1}^n \A_{i,j}$. For most~$\SN$, the frequency distribution $\rho(\degree)$ of the degree in $\G$ has been found by various researchers~\citep{barabasi99,albert02,barabasi03} to asymptotically follow the power law distribution of the form $\rho(\degree)=\alpha\times\degree^\power$. For social networks, and all other biological networks, the power usually takes the value $-3\le\power\le -2$. Having $\rho(\degree) \sim \alpha\times\degree^\power$ makes $\SN$ scale-free~\cite{albert02}. Figure~\ref{fig:author-author} shows an example of a small-world, scale-free graph that is naturally clustered and disconnected.

\subsection{Connected components}

The {\bf connected components} of an undirected graph~$\G$ are the maximal disjoint sets $V_1, V_2, \dots, V_n$ such that $\V=V_1\cup V_2\cup\cdots\cup V_n$, and the vertices $u,v\in V_i$ if and only if~$u$ is reachable from~$v$ and~$v$ is reachable from~$u$~\cite{woo89,cormen90}. Two methods are generally used to identify the connected components of~$\G$:  (1)~The breadth-first-search (BFS); and (2)~The depth-first-search (DFS). We can use any of these two to see if a certain path from~$u$ to~$v$ exists for each vertex pair of~$(u, v)$~\cite{ET75}. Given a starting vertex~$v_0$, BFS systematically search a given graph of vertices that has a path from~$v_0$. First, BFS lists all vertices that are adjacent to~$v_0$. Then, it starts again with another vertex~$v_i$ in the list that is directly connected with the previous vertex. The usual convention is to take the first vertex in the list as the~$v_i$. BFS again does the listing of vertices that are directly connected to~$v_i$. The algorithm stops when there are no more vertices that have a path from~$v_0$. Now, if there still exist vertices that are not listed after the BFS has been done, then the said graph is considered disconnected. The complexity of a BFS algorithm that returns all connected components is~$0(|\V|\times|\E|)$. 

In DFS, the traversal is done in a depth-first fashion, wherein the outcome is a forest of depth-first trees. Each tree in the forest contains vertices that belong to a different subgraph. The correctness of DFS as a test for graph connectivity follows directly from the definition of a spanning tree, and from the fact that the graph is undirected. This means that a depth-first tree is also a spanning tree of a graph induced by the set of vertices in the depth-first tree. Assuming that the graph is stored using a sparse representation, the run time of the DFS is~$\theta(|\E|)$.

\subsection{MCL}

The MCL starts from a random starting vertex~$v_0\in\G$ and walks to other vertices connected to~$v_0$. Here, $\G$ maybe described using a similarity matrix. The traversal usually does not leave the graph's cluster until many of the cluster's vertices have been visited. The idea of the algorithm is that it simulates ``flow'' within a graph. The flow is done iteratively wherein after each step, MCL demotes the edges within the distant nodes and promotes the edges of the nearby nodes. To do this, MCL takes the corresponding~$n\times n$ adjacency matrix~$\A$ of the graph~$\G$ and normalizes each column to obtain a stochastic matrix~$\M$. This includes adding the diagonal elements in the adjacency matrix to include self-loops for all nodes. After initializing the matrix, the algorithm uses two alternating functions: (1) expansion, which is used to flatten the stochastic distributions in the columns and causes the edges and paths of the random walker to become evenly spread; and (2) inflation, which contracts them to favor paths. It is said that the MCL algorithm's complexity is~$O(n^3)$, where~$n=|\V|$ is the number of vertices of the input graph. This is the same as the cost of multiplying two matrices of dimension~$n$. It is also noted that the inflation step of the algorithm has a complexity of~$O(n^2)$. The mathematical analysis on the time complexity of MCL is discussed in detail by~\citet{Dongen69}.

\subsection{Kamada-Kawai}

The KK algorithm~\cite{KK89} is commonly described as a ``spring-embedder'' where the vertices $v_1, v_2, \dots, v_n\in\V$ are considered particles that are mutually connected by springs in a dynamic system. Each vertex $v_i\in\V$ is initially located within the canvass with its two-dimensional coordinates $(x_i, y_i)$. The human-readable layout of vertices in the canvass is directly related to the dynamic balance of the energy~$E$ in the spring system. In other words, $E$~is modeled as a system of springs with a degree of elasticity wherein a desired resting length is achieved when the system reaches an equilibrium. This physical fact is described mathematically in Equation~\ref{eqn:energy}. The best layout for a given graph~$\G$ is at minimum~$E$.

\begin{eqnarray}
  E &=&  \frac{1}{2}\sum_{i=1}^{n-1} \sum_{j=i+1}^n k_{ij} \left(D_{i,j} - L\times d_{ij}\right)^2\label{eqn:energy}\\
  L &=& \frac{L_0}{\max_{i<j} d_{ij}}\label{eqn:L}\\
  k_{ij} &=& \frac{K}{d^2_{i,}}\label{eqn:kij}
\end{eqnarray}

In Equation~\ref{eqn:energy}, $D_{ij}=\sqrt{(x_i-x_j)^2 + (y_i-y_j)^2}$ is the Euclidean distance between vertices~$v_i$ and~$v_j$ while $d_{ij}$ is the graph theoretic shortest path~\cite{chartrand86}. $L$~is the desired length of the canvass edge. However, when the size of the canvass edge is already constrained, say as $L_0$, $L$~now (Equation~\ref{eqn:L}) depends on the graph theoretic diameter~\cite{chartrand86}, which is the distance between the farthest pair of vertices in a graph. The coefficient $k_{ij}$ (Equation~\ref{eqn:kij}) quantifies the strength of the spring that connects~$v_i$ and~$v_j$. In Equation~\ref{eqn:kij}, $K$~is a constant.
 
Given an initial location for each vertex, KK firstly calculates the ``energy'' or the sum of spring tension for each vertex. The initial vertex location is usually randomly assigned within the canvass. Some implementations of KK randomly initialize the vertices along the diameter of a circle. Whichever vertex initialization procedure is used, KK first finds the vertex~$v*$ with the highest energy. It then uses a modified Newton-Raphson procedure~\cite{rowe87} to compute the new positions of~$v*$ until the energy in the graph is minimum, or below a certain threshold~$\epsilon$. The necessary condition to find the minimum is,
$$
  \frac{\partial E}{\partial x_m} = \frac{\partial E}{\partial y_m} = 0, \quad\forall 1\le m \le |\V|.
$$
The above condition can be calculated by taking the derivatives of Equation~\ref{eqn:energy} with respect to~$x_m$ and~$y_m$:

\begin{eqnarray}
  \frac{\partial E}{\partial x_m} &=& \sum_{i\ne m} k_{mi}\left\{ (x_m - x_i) - \frac{l_{mi}(x_m-x_i)}{D_{mi}} \right\},\\
  \frac{\partial E}{\partial y_m} &=& \sum_{i\ne m} k_{mi}\left\{ (y_m - y_i) - \frac{l_{mi}(y_m-y_i)}{D_{mi}} \right\}.
\end{eqnarray}

After this, KK looks again for the vertex with the next highest energy and begins moving it. When all vertices have been moved, KK stops and the graph drawing is done. The complexity of the original KK algorithm is~$0(|\V|^3)$ which is just equivalent to finding the distances of all pairs of vertices in~$\G$ (i.e., the simple shortest-path algorithm of Floyd). After that, KK requires~$0(|\V|^2)$ to compute the Newton-Raphson iteration for all high-energy vertices. The reader is directed to the work of~\citet{KK89} for a thorough complexity analysis of KK.

Because of the ease of using the KK algorithm for drawing graphs, several modifications have been made to it. One of them is the modification for input graphs with non-uniformed vertex sizes. This modification uses an iterative KK (IKK) where a layout for a graph with arbitrarily sized-vertices is found by iteratively finding a nice layout of a similar graph with weighted edges and dimensionless vertices~\cite{HK02}.

\subsection{Drawing constraints}

The literature is not lacking on methodologies that allow one to visualize graph structures. In some of these methods, positioning of vertices are restricted to some location within the drawing canvass. For example, vertices could be located on grid points~\cite{batini86,TDB88}, within concentric circles~\cite{carpano80}, or along parallel lines~\cite{STT81,rowe87}. Edges, on the other hand, maybe drawn as straight lines, polygonal lines, or curves. In our drawings, we did not put a constraint on the location of the vertices, while we have drawn the edges as straight lines. The main task of our algorithm, is therefore, to find a location for the vertices of a given graph such that the number of edge crossings is minimized, and at the same time, uniformly distribute the vertices and the edges within the canvass for easier readability by humans.

\section{Hybrid drawing}

\begin{figure}[htb]
\centering\epsfig{file=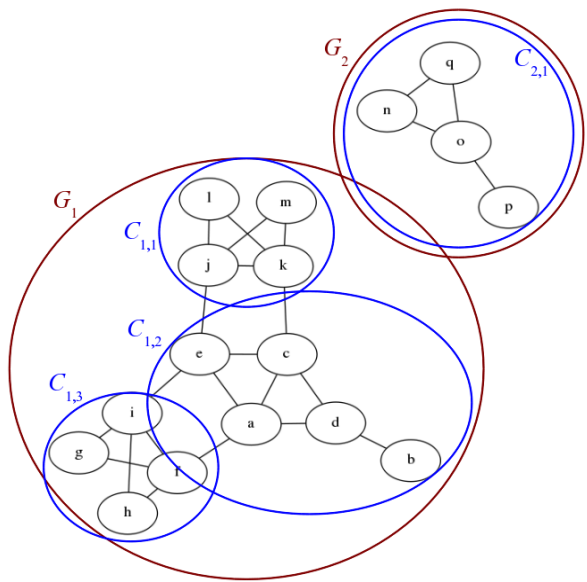, width=3in}
\caption{An example graph~$\G$ with two unconnected subgraphs $G_1$ and $G_2$ (colored in the digital format of this paper). The red circles identify the two subgraphs $G_1$ and $G_2$ found by the first step. The blue circles identify the four clusters $C_{1,1}, C_{1,2}, C_{1,2}$ and~$C_{2,1}$ found by the second step. There are three clusters in $G_1$ and only one in $G_2$. The nodes in each cluster were drawn in step~3. Clusters $C_{1,1}$, $C_{1,2}$ and $C_{1,3}$ were considered as ``phantom vertices'' and were arranged using IKK in steps~4 and~5. Subgraphs $G_1$ and $G_2$ were considered as ``phantom vertices'' and were arranged using IKK in steps~6 and~7.} \label{fig:sample}. 
\end{figure}

We discuss the procedure for our HGDA using the graph shown in Figure~\ref{fig:sample} as an illustrative example. The procedure  is as follows:
\begin{enumerate}
\item On an input $\G(\V,\E)$, run DFS to output $n$~subgraphs $\{G_1(V_1,E_1), G_2(V_2,E_2), \dots, G_n(V_n,E_n)\}$. Here, $\V=V_1 \cup V_2 \cup \cdots \cup V_n$, $\E=E_1 \cup E_2 \cup \cdots \cup E_n$, and $n\le |V|$. As discussed above, this step has a complexity of $\theta(|\E|)$.
\item For each sub-graph~$G_i$, run MCL to find the clusters in each~$G_i$, $\forall i=1,2,\dots,n$. The output of the $i$th MCL is $m$~clusters $\{C_{i,1}, C_{i,2},\dots,C_{i,m}\}$. Each cluster $C_{i,j}$ has an associated set of vertices $v_{i,j}$. Here, $V_i = v_{i,1} \cup v_{i,2} \cup \cdots\cup v_{i,m}$ and $m\le |V_i|$. This step has a complexity of $O(n \times (\max_{i=1}^n(|V_i|))^3)$.
\item For each cluster~$C_{i,j}$, run KK to rearrange the vertices in $v_{i,j}$. This step has a time complexity of $O(m \times n \times (\max(|v_{1,1}|, \dots, |v_{m,n}|))^3)$. 
\item\label{step:4} Consider each cluster~$C_{i,j}$ as one big ``phantom vertex'' for a temporary subgraph~$G'_i$. If there is at least one edge going from one vertex in the current cluster~$C_{i,j}$ to another vertex in another cluster~$C_{i,k}$, create a ``phantom edge'' connecting~$C_{i,j}$ and~$C_{i,k}$. The complexity of this step is $O(n\times m)$ to connect the $m$~phantom vertices with $m-1$~phantom edges.
\item Run IKK on~$G'_i$ to rearrange the clusters within the sub-graph~$G_i$. It should be noted that because clusters are now considered as a vertex for the sub-graph~$G_i$, the phantom vertex has already gained its own size. Because of this, IKK is useful because of its power in drawing nice layouts for graphs with vertices that have different sizes. The complexity of this step is $O(n \times m^3)$ because there are only $n$~subgraphs with $m$~phantom vertices each. 
\item Consider each~$G_i$ as one phantom vertex for a temporary graph~$G*$. Since all subgraphs are disconnected from each other, make each~$G_i$ be connected to at most 2 other phantom vertices only and no two phantom vertices have at least one same vertex connected into it to avoid creating a cyclic graph. As in step~\ref{step:4}, the complexity of this step is $O(n)$ to connect the $n$~phantom vertices with $n-1$~phantom edges.
\item Run IKK on~$G*$ to rearrange the sub-graphs. Again, using IKK is useful here because sub-graphs, which are now considered as phantom vertices, will be of different sizes and has dimensions. The complexity of this step is $O(n^3)$ because there are only $n$~phantom vertices corresponding to $n$~subgraphs. 
\end{enumerate}

\subsection{Fully-connected graphs as input}

On an input of a fully-connected graph $\G(\V,\E)$, HGDA will have a complexity of $\theta(|\V|\times (|\V|-1)/2) = O(|\V|^2)$ in step one. Step two, however, will have $O(|\V|^3)$ since there is only one subgraph and the lone subgraph has $|\V|$~vertices. Since $\G$ is fully connected, only one cluster will be created from MCL and thus step three will have a time complexity of $O(|\V|^3)$. Each remaining steps will only run in $\theta(1)$ because the number of clusters found is one, while the number of subgraphs created is also one. Thus, for a fully-connected~$\G$ as an input, HGDA will run in $O(|\V|^2 + O(|\V|^3) + O(|\V|^3) + \theta(1) + \theta(1) = O(|\V|^3)$. 

\subsection{Graphs with $\E=\emptyset$ as input}

On an input of a graph~$\G(\V,\emptyset)$ with no edge, this means that there are $|\V|$~sub-graphs, each with only one vertex. Step one will have a zero time complexity. However, step two is exactly $\theta(|\V|)$, while step three is $O(1)$. Steps five and seven will run $O(|\V|)$ and $O(|\V|^3)$, respectively. Thus, for an input of $\G(\V,\emptyset)$, HGDA will run in $\theta(|\V|) + O(1) + O(|\V|) + O(|\V|^3) = O(|\V|^3)$.

\section{Parallel Implementation}

Our proposed HGDA needs to be run on parallel processors in order to efficiently draw large, naturally-clustered, disconnected graphs. In this section, we present our implementation of the HGDA over a parallel random access machine (PRAM) architecture and derive the corresponding parallel complexities per step. We assume here that our PRAM has $p$~processing units (PUs) that can compute in parallel.

\subsection{Parallel DFS}

The search for connected components of the input graph~$\G$ can be parallelized by partitioning the adjacency matrix~$\A$ into $p$~parts and then assigning each part to one of $p$~PUs. Each PU~$\PU_i$ has an associated subgraph $G_i$ of $\G$, where $G_i(\V,E_i)$ and $E_i$ are the set of edges that correspond to the portion of~$\A$ assigned to~$\PU_i$. We implement the following steps:
\begin{enumerate}
\item\label{DFS-step:1} Each~$\PU_i$ computes the depth-first spanning forest of~$G_i$ to construct $p$~spanning forests;
\item\label{DFS-step:2} We merge the spanning forests pairwise until only one spanning forest remains.
\end{enumerate} 
The remaining spanning forest has the property that two vertices~$u$ and~$v$ are in the same connected component if they are in the same tree. Step~\ref{DFS-step:1} above can be computed sequentially by using any of the~\citet{kruskal56},~\citet{prim57}, or~\citet{sollin77} algorithms. However, a parallel algorithm exists that uses $p=|\V|^2$ on a concurrent-read, exclusive-write PRAM to solve the problem in time $\Theta(\log^2 |\V|)$~\cite{savage81}. To implement step~\ref{DFS-step:2} above efficiently, our parallel implementation uses disjoint sets of edges. Assume that each tree~$t_{i,j}$ in the spanning forest~$T_i$ of a subgraph~$G_i$ of~$\G$ is represented by a set. The sets for different trees are pairwise disjoint. We defined the following functions to be applied on the disjoint sets:
\begin{description}
\item[{\bf find($x$)}:] This function returns a pointer to the representative element of the set containing~$x$, where each set has its own unique representative.
\item[{\bf union($x$, $y$)}:] This function unites the sets containing the elements~$x$ and~$y$. The two sets are disjoint prior to the operation.
\end{description}
Let $T_i$ and $T_j$ be the two spanning forests to be merged. We merge the spanning forest as follows. At most $|\V|-1$ edges of one forest are merged with the edges of another forest. For each edge $(u,v)\in T_i$, a {\bf find} operation is performed for each vertex to determine if the two vertices are already in the same tree of~$T_j$. If not, then the two trees of~$T_j$ containing~$u$ and~$v$ are united by the {\bf union} function. We can see here that merging~$T_i$ and~$T_j$ requires at most $2(|\V|-1)$ {\bf find} calls and $(|\V|-1)$ {\bf union} calls. Thus, the cost of merging is~$O(|\V|)$. The parallel DFS has a parallel complexity of $\theta((\log^2 |\V|)$ because it is dominated by step~\ref{DFS-step:1} above.

\subsection{Parallel MCL}

Since MCL is based on the simulation of stochastic ``current'' flow in graphs, an analytical method can not be performed for implementing the parallel MCL over PRAM. However, several implementations, such as those by~\citet{olman09} and~\citet{bustamam09}, have been performed over a message-passing architecture wherein the respective runtimes were experimentally determined.

\section{Conclusion}

We developed a hybrid graph drawing algorithm by incorporating in series:
\begin{enumerate}
\item the DFS to find the $n$~connected components $G_i$ ($\forall i=1,\dots,n$) of an input graph~$\G$,
\item the MCL to find the $m$~clusters of vertices in each connected component,
\item the KK to layout the vertices in $j$th cluster,
\item the IKK to layout the clusters as phantom vertices, and
\item another IKK to layout the components as another phantom vertices.
 \end{enumerate}
We derived the runtime complexity of our hybrid algorithm by considering input graphs in extreme cases: a fully connected graph $\G(\V,\E)$ and a graph with no edge $\G(\V,\emptyset)$. With these input graphs, we found out that HGDA has $O(|\V|^3)$ runtime complexity, where $\V$~is the set of vertices of the input graph~$\G$. Although we found that the runtime of HGDA is slower than that of the KK or IKK, our purpose here is not to improve the runtime of the drawing algorithm, but instead to ``nicely'' draw large, naturally-clustered, and disconnected graphs that usually model the complex inter-relationships among entities in social, biological, natural, and artificial networks. We designed an implementation of parallel DFS over a PRAM and found its parallel runtime to be $\Theta(\log^2|\V|)$ if $p=|\V|^2$.

\section{Acknowledgments}

This research effort is funded by the Institute of Computer Science through UPLBGF \#2326103 and UPLBFI \#2004987. We also thank the nameless reviewers whose suggestions vastly contributed to the improvement of the paper.

\bibliographystyle{unsrtnat}
\bibliography{graph-drawing}

\end{document}